\documentstyle[prl,multicol,aps,epsfig]{revtex}
\begin{document}

\title{Reply to Suslov}

\author{Keith Slevin}
\address{Dept. of Physics, Graduate School of Science,
Osaka University, \\ 1-1 Machikaneyama, Toyonaka,
Osaka 560-0043, Japan}

\author{Tomi Ohtsuki}
\address{Department of Physics, Sophia University,
Kioi-cho 7-1, Chiyoda-ku, Tokyo 102-8554, Japan}

\maketitle

\begin{abstract}
In a recent submission to this archive \cite{Suslov}, Suslov has claimed that our recent
 numerical estimate of the critical exponent of the Anderson transition
$\nu=1.57\pm.02$ is in error and that the available numerical data are 
consistent with a value of $\nu=1$. We contest this claim and demonstrate using
Suslov's own scaling procedure that $\nu\ne 1$.

\end{abstract}

\begin{multicols}{2}

\section{Our published estimate}

In \cite{SO99} we published an estimate of the critical exponent of the Anderson
transition based on a finite size scaling analysis of the localisation length
$\lambda$  of electrons on quasi-1d bars of dimensions $L\times L\times L_z$. In
our published work $L$ ranged from $L=4$ to $L=14$ and $L_z$ was of the order
of $10^7$ in units of the lattice spacing of the tight binding model which was
studied. In more recent numerical studies we have extended the maximum system
size for which data is available to $L=18$.

The critical exponent is estimated by fitting data for the localisation as a function
of system size $L$ and the magnitude of the random potential fluctuations measured
by a disorder parameter $W$ to a finite size scaling form that contains one relevant 
scaling variable $\chi$ and one irrelevant scaling variable $\psi$,
\begin{equation}
\Lambda = F_0 \left(\chi L^{1/\nu} \right) + \psi L^y F_1\left(\chi L^{1/\nu} \right)
\end{equation}
The functions $F_0$ are $F_1$ are expanded in Taylor series, as are both
scaling variables.
An important point to note is that the relevant variable is zero at the
critical disorder $W_c$, so that,
\begin{equation}
\chi= b_1 \tau + b_2 \tau^2 + \ldots
\label{chi}
\end{equation}
while the irrelevant variable is not, so that
\begin{equation}
\psi=c_0 + c_1 \tau + \ldots \ldots
\label{psi}
\end{equation}
where 
\begin{equation}
\tau=\frac{W_c-W}{W_c}
\end{equation}
In practice, when fitting our data for the box distributed random potential
for example, we
truncate the expansion for $\chi$ at second order, 
the expansion for $\psi$ at zero order and the expansions for 
$F_0$ and $F_1$ at third order.

Using these equations to fit our numerical data we estimated $\nu=1.57\pm.02$,
where the error is a $95\%$ confidence interval, for a box distributed random potential.
We obtained estimates consistent with this for Gaussian and Lorentzian distributed 
random potentials.
This value of the exponent is also consistent with a completely 
independent estimate based on the scaling of the conductance of $L\times L \times L$ 
three dimensional systems for $L\le 16$ \cite{SO01} and 
with the published estimates of other authors
\cite{MacK94,Cain,Markos,MildePRB,MildeEJP}.

\section{The validity of confidence intervals}

The fitting of a set of parameters $X$ given a set of data $D$ and a 
model $M$ is often based on finding the maximum of the probability
\begin{equation}
p(D|X,M)
\label{prob}
\end{equation}
This is called a maximum likelihood method since the parameters are
chosen so as to make the probability of the observed data a maximum.
If the probability is calculated by assuming that differences between
the observed and predicted data are due to random measurement
errors which are independently distributed according to a Gaussian
distribution we are led to the well known $\chi^2$ fitting procedure.

It is crucial to note that analysis is predicated on the assumption
that the model is correct. In particular, the estimate of the accuracy 
of the fitted parameters are all contingent on this assumption.
To be sure, if the goodness of fit is low, then one can doubt the model.
However, a reasonable goodness of fit is not a guarantee that the model
is correct.

What then are the assumptions underlying our model? We take for granted
that the Anderson transition is a continuous phase transition which
can be described by the renormalisation group. In addition we assume
that the deviations from scaling in our numerical data can be described in 
practice by a single irrelevant variable. Given these assumptions we arrive
at the accuracy quoted for the exponent.

We would like to emphasise that the procedure for arriving at the
accuracy of the exponent is a rational and rigorous procedure and the
results have a precise scientific meaning and are not a matter of opinion.
As with any mathematical procedure the validity of the results depends
on the correctness of the assumptions made. These have been clearly
stated. 

It is of course always possible that if data for much larger
system sizes were available some limitation of the model we use to fit
the data might be exposed. If this turns out to be the case then our estimate of 
the exponent might have to be revised. Nevertheless, we strongly object to
Suslov's characterisation of our results as ^^ ^^ evident
disinformation." They are nothing of the sort.

We think it worth pointing out that with the methods currently used for
estimating the exponent the computational time increases as $L^7$. It
seems unlikely that accurate data for substantially larger systems
will be available any time soon. In this context we are forced to do
the best we can with the available data by proceeding in a logical
manner. In our opinion the libelous and ill informed comments in Ref. 
\cite{Suslov} contribute nothing.

\section{Spin glass, what spin glass?}

Suslov claims that out fit is highly non- linear and that maximising
the probability (\ref{prob}) is somehow akin to finding the 
potential minimum in a spin glass. 
Though our model does contain many parameters many of them are linear.
For example, of the 12 parameters required to fit the data for the box 
distributed random potential in Ref. \cite{SO99} 7 are linear. 
Hardly a spin glass ...

\section{An error in Suslov's paper}

Suslov proposes what he calls ^^ ^^ a simple procedure to deal with
corrections to scaling."
Unfortunately, his proposal is flawed because of an elementary error
in Eq. (6) of his paper.
Suslov assumes incorrectly that the irrelevant scaling variable is 
zero at $W=W_c$ or $\tau=0$. In fact, only the relevant variable
changes sign at $W_c$ as is clear in the correct expansions 
(\ref{chi}) and (\ref{psi}) given above.
If we look at Eq (9) of Suslov's paper we see that the quantity
being fitted must be independent of $L$ at $W=W_c$.
However, the most important correction to scaling which is present
in our numerical data is precisely a size dependence of $\Lambda$
at $W_c$. Thus Suslov's method is a non- starter as far as modeling
our data over the full range of systems sizes is concerned. To use Suslov's 
method we are forced to discard
any data for which corrections to scaling due to an irrelevant variable
are statistically significant.

\section{Suslov's procedure with our data}

By restricting the ranges of system sizes and disorder, we have been able 
to fit some of our numerical data for the box
distributed random potential to the following
form based on the suggestion in Suslov's paper.
\begin{equation}
\Lambda = a_0 + \tau f(L) + a_2 \left(\tau f(L) \right)^2
\label{suslovfit}
\end{equation}
The fitting parameters are $W_c$, $a_0$, $a_2$ and one
$f_i\equiv f\left(L_i \right)$ for each system size present in
the data.
In this case this is a total of 8 parameters, 6 of which are
non- linear.
The results are shown in Figure \ref{F1}. The critical
exponent is then estimated by fitting the $f_i$ versus
$L_i$ to a power law which introduces a further two fitting
parameters, giving 10 in total.
The result is $\nu=1.53(46,60)$ and is shown in Figure \ref{F2}.
The numbers in brackets give the $95\%$ confidence interval. For 
comparison we also plot a line corresponding to $\nu=1$.
We leave it to the reader to speculate on whether or not he or she 
thinks $\nu=1$ might be recovered for much larger systems sizes.
We do not see any evidence for such a claim.

\section{Suslov versus Slevin- Ohtsuki fitting}

When we compare the two fitting procedures the following points
present themselves.
First, Suslov's method cannot describe the most important corrections
to scaling which are present in the numerical data. Second, given that
one non- linear parameter is needed for each system size, we end up
with more, not fewer, non- linear parameters than the fit we used in
Ref. \cite{SO99}.

\section{Final thoughts}

In conclusion, Suslov's fitting scheme [1] does not correctly
take into account the most important corrections to scaling in
our data.  If we restrict ourselves to the larger systems simulated
where corrections to scaling are negligible and apply the fitting 
scheme suggested in Ref. [1], we find $\nu =1.53\pm.07$, consistent 
with our previous estimate in [2] but not with $\nu =1$.

Someone once said that the only certainties in Life are Death 
and Taxes.
Certainly current estimates of the critical exponent of the Anderson transition
have not reached this level of certainty and it is important to keep an
open mind about how the estimates might need to be revised when new
data becomes available.
Yet we feel there is no need to be overly pessimistic concerning the accuracy
of our current estimates.

\end{multicols}

\newpage

\begin{center}

\begin{figure}
\epsfig{file=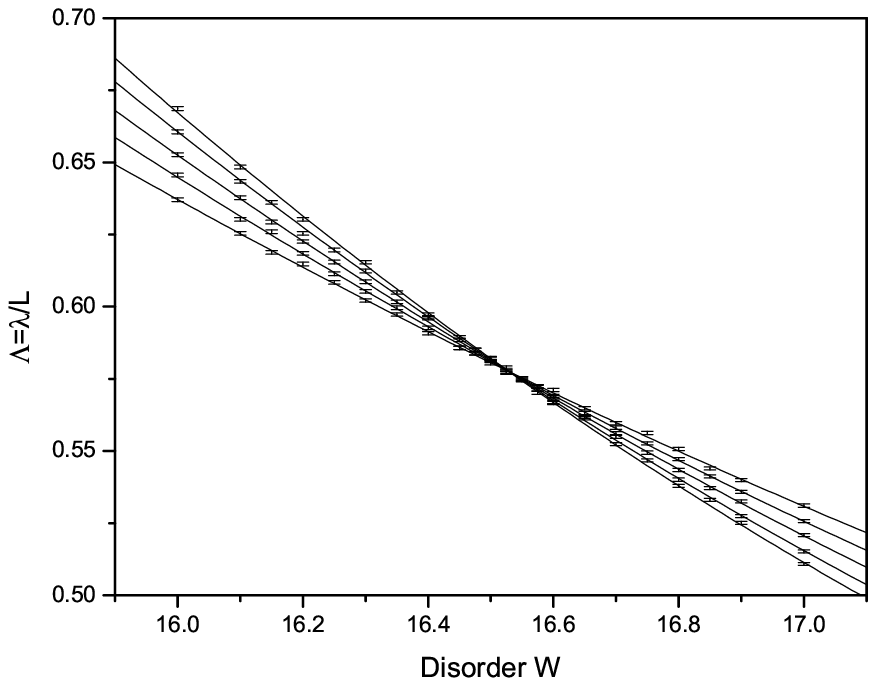,width=12.5cm}
\caption{A fit of our numerical data for the box
distributed random potential with disorder $16\le W\le 17$ and
$10\le L\le 18$ to Eq. (\ref{suslovfit}).}
\label{F1}
\end{figure}

\begin{figure}
\epsfig{file=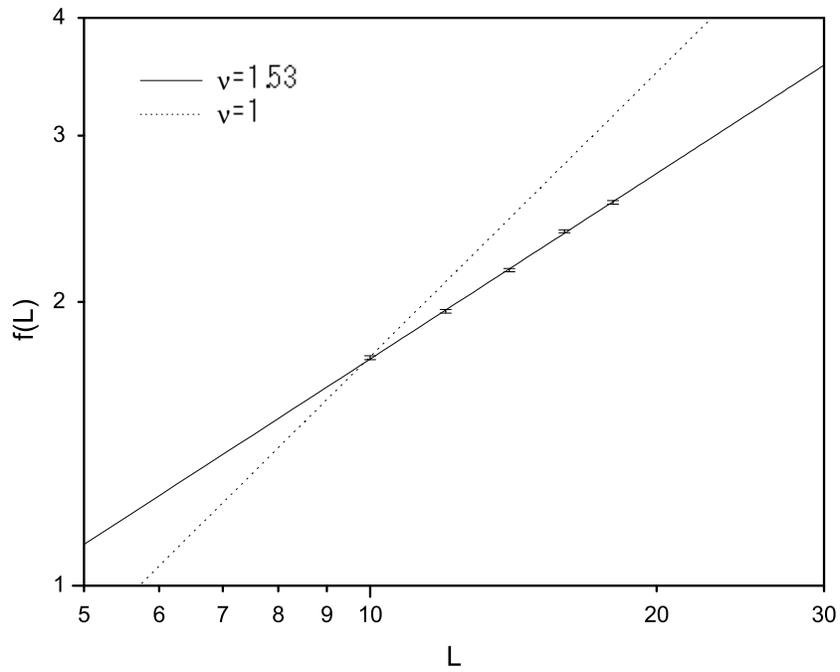,width=12.5cm}
\caption{The system size dependence of the fitting parameters
$f(L)$ from which the critical exponent is estimated. The best fit
$\nu=1.53$ and for comparison a slope corresponding to $\nu=1$ are shown.}
\label{F2}
\end{figure}

\end{center}

\end{document}